\documentclass[aps,prd,preprintnumbers,showpacs]{revtex4}
\usepackage[dvips]{graphicx}
\begin{document}

%
%

\eprint{Nisho-3-2017}
\title{QCD monopole and sigma meson coupling}
\author{Aiichi Iwazaki}
\affiliation{International Economics and Politics, Nishogakusha University,\\ 
6-16 3-bantyo Chiyoda Tokyo 102-8336, Japan.}   
\date{Feb. 30, 2017}
\begin{abstract}
Under the assumption of the Abelian dominance in QCD, 
we show that chiral condensate is locally present around a QCD monopole.
The appearance of the chiral condensate around a GUT monopole
was shown in the previous analysis of the Rubakov effect.
We apply a similar analysis to the QCD monopole. 
It follows that
the condensation of the monopole carrying the chiral condensate leads to the chiral symmetry breaking as well as
quark confinement.
To realize the result explicitly, we present a phenomenological linear sigma model coupled with the monopoles,
in which the monopole condensation causes the chiral symmetry breaking as well as confinement.
The monopoles are assumed to be described by a model of dual superconductor.
Because the monopoles couple with mesons, we point out the presence of an observable color singlet monopole
coupled with the mesons.  
\end{abstract}
\hspace*{0.3cm}
\pacs{12.38.Aw,11.30.Rd,12.38.Lg,12.39.Fe}

\hspace*{1cm}

\maketitle

It has been shown with lattice gauge theories\cite{karsch} that quark confinement and chiral symmetry breaking simultaneously arises in SU(3) gauge theory
with massless quark color triplets.
That is, the transition temperature between confinement and deconfinement phases 
almost coincides with the transition temperature 
between chiral symmetric and antisymmetric phases. Although extensive studies\cite{suga,miyamura,w,fuku,gat,sch,iritani} have been performed,
the explicit connection between the confinement and the chiral symmetry breaking has been still not
clear.  
The confinement
is caused by the monopole condensation\cite{nambu,man,thooft} in the analysis with the use of maximal Abelian gauge\cite{mag}. 
On the other hand, the chiral symmetry
breaking is caused by the chiral condensation of quark-antiquark pair. It arises with 
instanton effects through chiral anomaly. No theoretical relation between
the monopole condensate and the chiral condensate is found, although there were 
numerical evidences\cite{miyamura} that the monopole condensates are
correlated with the chiral condensates.

In this paper we show that the chiral condensate is locally present around a QCD monopole. Thus,
the monopole condensation leads to the chiral symmetry breaking. 
As we briefly explain below,  when a massless fermion collides with a monopole,
the fermion flips its chirality; the chirality is not conserved around a monopole. 
The flip of the chirality can arise owing to the presence of the local chiral condensate around a monopole.
The fact was shown in previous analyses\cite{rubakov,callan,ezawa} of Rubakov effect where massless fermions scattering with a monopole 
in grand unified theories ( GUT monopole ) was explored. 
Although the main purpose of the analyses
was to show the presence of the Rubakov effect, that is, the baryon decay, 
the presence of the chiral condensate was also shown as a by-product. 

The GUT monopoles discussed in the Rubakov effect are ones arising in grand unified theories
in which SU(2) gauge subgroup is broken into U(1) gauge group by a Higgs triplet. They are 'tHooft-Polyakov monopoles\cite{mono}.
The effect was studied in SU(2) gauge theory with the monopole excitations. ( The monopoles result from the spontaneous symmetry breaking of SU(2)
to U(1) with the Higgs triplet. ) 
The relevant fields for the effect are the U(1) gauge fields and massless fermions doublet ( quarks and leptons ); massive gauge bosons and Higgs fields are irrelevant.
( The fermions do not couple with the triplet Higgs. Thus, they are massless and there are no zero modes associated with the  'tHooft-Polyakov monopole. 
The presence of the chiral condensate shown in this paper is not related with the zero modes. ) 
On the other hand,
the QCD monopoles are those arising in QCD when we assume Abelian dominance\cite{iwa,suzuki}. 
It dictates that only maximal Abelian gauge fields coupled with QCD monopoles
are massless and relevant for the analysis of low energy properties in QCD. 
It apparently seems that the SU(3) gauge symmetry is broken into the maximal Abelian symmetry,
although it is not broken in reality.  
Thus, the situation is similar to the case in the Rubakov effect in which SU(2) gauge group is broken into U(1) gauge group.
The U(1) gauge fields in the Rubakov effect correspond to the Maximal Abelian gauge fields of SU(3)
and the 'tHooft-Polyakov monopoles ( GUT monopoles )
correspond to
Wu-Yang monopoles ( QCD monopoles ) under the assumption of the Abelian dominance. The massless fermions are u, d quarks in QCD. Then, it follows that
the chiral condensate $\langle \bar{q}{q} \rangle \neq 0$ of the quarks $q$ is locally present around the QCD monopole. 
Therefore, it is obvious that the monopole condensation in QCD vacuum causes the chiral symmetry breaking as well as the confinement of quarks and gluons.

\vspace{0.1cm}
To realize the result explicitly, 
we present a phenomenological sigma  model 
coupled with the monopoles. The monopoles are 
supposed to be described in a model of dual superconductor\cite{nambu,man,dual,maedan} 
in which quark confinement is effectively realized. 
Thus, our model is composed of a linear sigma model with the symmetry $\rm SU(2)_R\times SU(2)_L$ and a model of dual superconductor.
The model involves an interaction between the monopole and the sigma meson so that 
the chiral symmetry breaking takes place as a result of the monopole condensation. 
Namely, the sigma field acquires nonzero vacuum expectation value owing to the
presence of the monopole condensate. On the other hand,
the chiral symmetry breaking in the standard sigma models happens owing to the presence
of the imaginary mass of the sigma field assumed ad hoc. 
( In our model we do not assume the presence of such an imaginary mass. )
Accordingly,
the QCD monopoles in our model can be observable owing to the interaction with the mesons of the sigma model.

As we show below, the Weyl symmetry\cite{ichie} in the dual superconducting model, which is a 
discrete symmetry of color SU(3), 
plays an important role in restricting observable particles. 
The dual gauge fields are not observable, because they are not invariant
under the transformation of the Weyl symmetry.
The three types of the monopole are also not observable in the phase without monopole condensation.
The Weyl symmetry requires that the observable ones are magnetically neutral composite of the monopoles.
On the other hand, one of them is observable in the phase with the monopole condensation.
This is because  it becomes neutral in the phase with the monopole condensation.
Because the monopole is a glue ball,
we are tempted to identify the monopole with actually observed $f_0$ meson.

\vspace{0.2cm}
Now, we briefly explain how the chiral condensate arises around a magnetic monopole.
For simplicity we discuss a massless fermion $\Psi$ doublet around a monopole in SU(2) gauge theory,

\begin{equation}
\label{1}
\gamma_{\mu}(i\partial^{\mu}-\frac{e}{2}A^{\mu})\Psi=0
\end{equation}
where the gauge potentials $A^{\mu}=A^{\mu}_a\tau_a/2 $ denotes a field configuration of a monopole
coupled with the fermion doublet ( $\tau_a$ denotes Pauli matrices ), for example 

\begin{equation}
A^{0}_a=0, \quad A^{i}_a=\epsilon_{a i j}\frac{x^j F(r)}{e r^2}.
\end{equation}
with charge indices $a=1,2,3$ and radial coordinate $r$. 

When $F\equiv1$, $A^{\mu}_a$ represents Wu-Yang monopole, a solution of Yang-Mills equations, which is singular at the origin $r=0$.
On the other hand, when the gauge group SU(2) is 
broken into the U(1) gauge group with a Higgs triplet,
a monopole solution is present with the smooth function $F(r)$;
$F(r)=1$ for $r>r_c$ and $F(r)\simeq 0$ for $r<r_c$. 
The core of the monopole is in the region $r<r_c$.
The solution is
regular at $r=0$ because $F(r=0)=0$. 
The monopole solution is called as 'tHooft-Polyakov monopole
relevant to the Rubakov effect. ( We assume that the fermion doublet does not couple with the triplet Higgs field. )
Apart from the monopole core, the field configurations of both monopoles are
identical; $F(r)=1$ at $r >r_c $ ( $r_c$ is extremely small compared with hadronic scale. )
Thus, as long as we consider low energy phenomena, there are no distinctions between the GUT monopole and the QCD monopole.
The Dirac equation (\ref{1}) holds both for the 'tHooft-Polyakov monopole and the Wu-Yang monopole 
when the fermions are outside the monopole core  $r>r_c$.
However, because S wave fermions approach the core,
the regularity of the GUT monopole solution imposes a boundary condition at $r=0$ 
for the fermions, as we show below. 

As is well known, the fermions around the monopole has the conserved angular momentum, $\vec{J}=\vec{L}+\vec{S}+\vec{\tau}/2$
where $\vec{L} \,( \vec{S}=\vec{\sigma}/2 ) $ represents orbital (spin) angular momentum and  $\vec{\tau}/2$ represents
an extra angular momentum of the fermion doublet associated with the monopole ( $\sigma_a$ denote Pauli matrices ).
Because the only fermion with orbital angular momentum $\vec{L}=0$ can be close to the monopole,
we only consider the component $\vec{J}=\vec{L}=0$.
We solve the equation (\ref{1}) for $r>0$ using the unitary gauge in which the fermions are composed of the charge eigenstates $\Psi_{\pm}$ such as $\Psi=(\Psi_+,\Psi_-)$
and the monopole $A_a^{\mu}$ with $F=1$ are transformed into the Dirac monopole.
The charge is defined as $\tau_3/2$.
Then, the components with $\vec{J}=\vec{L}=0$ are given by\cite{ezawa}

\begin{equation}
\Psi_{\pm}=\frac{1}{r}\left(\begin{array}{l}f_{\pm} (r,t) \\ \mp ig_{\pm}(r,t)\end{array}\right )\eta_{\pm} \quad \mbox{with} \quad
\frac{\sigma_i x_i}{r}\eta_{\pm}=\pm \eta_{\pm}.
\end{equation}

Then the equation is decomposed into two independent equations,

\begin{equation}
i\bar{\gamma}_{\nu}\partial^{\nu}\psi_{\pm}=0 \quad \mbox{with} \quad \psi_{\pm}\equiv\left(\begin{array}{l}f_{\pm} (r,t) \\  -ig_{\pm}(r,t)\end{array}\right) 
\end{equation}
with $\nu=0,1$, $x_0=t$ and $x_1=r$, where two dimensional gamma matrices are defined by

\begin{equation}
\bar{\gamma}^0=\left(\begin{array}{rr} 1 & 0 \\ 0  &-1 \\ \end{array}\right), \quad \bar{\gamma}^1=\left(\begin{array}{rr} 0 & 1 \\ -1  & 0 \\ \end{array}\right).
\end{equation}
We can easily solve the two dimensional equations. The solutions are characterized by their chiralities and charges as well as  
their motions i.e. incoming or outgoing.

We should make a comment that the right ( left ) handed projection operator for $\psi_{+}$ is given  such that 
$\frac{1}{2}(1+\bar{\gamma}_5)$ ( $\frac{1}{2}(1-\bar{\gamma}_5)$ ) for $\psi_{+}$, while they are given such as 
$\frac{1}{2}(1-\bar{\gamma}_5)$ (  $\frac{1}{2}(1+\bar{\gamma}_5) $ )  for $\psi_{-}$
with $\bar{\gamma}_5=\left(\begin{array}{rr} 0 & 1 \\ 1  &0 \\ \end{array}\right)$.
Therefore, when $E>0$, the solution $\psi_{+} =\exp(-iE(t-r))\left(\begin{array}{l}1 \\ 1\end{array}\right )$
 describes outgoing positive charge and right handed particles, while the solution  $\psi_{-} =\exp(-iE(t-r))\left(\begin{array}{l}1 \\ 1\end{array}\right )$
describes outgoing negative charge and left handed particles. Similarly, the solution $\psi_{+} =\exp(-iE(t+r))\left(\begin{array}{l}1 \\ -1\end{array}\right )$
describes incoming positive charge and left handed particles, while the solution  $\psi_{-} =\exp(-iE(t+r))\left(\begin{array}{l}1 \\ -1\end{array}\right )$
does incoming negative charge and right handed particles.

Thus, when an incoming fermion with positive charge is left handed, after the scattering with the monopole,
the outgoing fermion with the positive charge must be right handed. The chirality changes. If the chirality is conserved,
the outgoing left handed fermion must carry negative charge.  The charge is not conserved.   
Either of the charge or the chirality is not conserved.
The circumstance can be understood easily by seeing the conservation of the angular momentum $\vec{J}$,
that is, the conservation of the quantity $\vec{J}\cdot\vec{n}=\vec{S}\cdot\vec{n}+\vec{\tau}\cdot \vec{n}$ with $\vec{n}\equiv \vec{r}/r $.
For instance, the change of the chirality leads to the conservation of the spin direction, $\vec{S}\to \vec{S}$.
Thus, the electric charge do not changes; $\vec{\tau}\to \vec{\tau}$. On the other hand, the conservation of the chirality leads to
the change of the spin direction,  $\vec{S}\to -\vec{S}$. Thus, the charge is not conserved;
$\vec{\tau}\to -\vec{\tau}$.

\vspace{0.1cm}
 Obviously these solutions hold only at $r>0$ especially for the Wu-Yang monopole
because the monopole configuration is not well defined at $r=0$. 
It was pointed out\cite{boundary} that we need to impose
a boundary condition at $r=0$ for the monopole-fermion system being well defined. 
The incoming fermions should go out from the monopole after the scattering, preserving both their chiralities and charges 
since the chiralities and charges are conserved when there are no boundary conditions at $r=0$. But there are no such states as explained above. That is,
the outgoing fermions must carry different quantum numbers from those the incoming fermions carry. 
For the monopole-fermion system to be well defined, we need a boundary condition which mixes 
the fermions with different quantum numbers.  
The boundary conditions we impose may cause chirality non conservation or charge non conservation. 
For example, $\psi_{+}(r=0)=\bar{\gamma}_0\psi_{-}(r=0)$ or $\psi_{+}(r=0)=\bar{\gamma}_0\psi_{+}(r=0)$.
The first one leads to the charge non conservation, while the second one does to the chirality non conservation.
In the analysis of the Rubakov effect, we are automatically led\cite{rubakov} to take the first one,
that is, the charged non conserved but chirality conserved ( and baryon number non conserved ) boundary condition.
It is derived by solving the Dirac equation (\ref{1}) by taking account of the regularity $F(r=0)=0$ of the 'tHooft-Polyakov monopole, that is,
$i\bar{\gamma}_{\nu}\partial^{\nu}\eta+(\frac{1-F}{r})\eta=0$ with $\eta\equiv (\psi_{+}-\bar{\gamma}_0\psi_{-})$ where
the last term vanishes outside the core of the monopole but it remains inside the core.
Thus, we need to take the boundary condition  $\psi_{+}-\bar{\gamma}_0\psi_{-}=0$ at $r=0$.

The boundary condition is for the fermions scattered by the 'tHooft-Polyakov monopoles.
On the other hand,  
it is not clear in QCD which boundary conditions should be chosen. 
Massless quarks in QCD are scattered by the Wu-Yang monopole which is singular at $r=0$. 
But even if we choose any boundary conditions in QCD, the chiral condensate locally arises around a QCD monopole,
as we explain below.

\vspace{0.1cm}
Now, 
after briefly explaining the presence of the chiral condensate around the GUT monopole,
we proceed to explain how the chiral condensates arise locally around the QDC monopole.
The relevant fields to the Rubakov effect are U(1) gauge fields and charged massless fermion doublet coupled with
the GUT monopole. 
Heavy gauge bosons and Higgs fields
are irrelevant for the analysis, 
because only processes with much lower energies than the masses of the heavy bosons are considered.
The fermions colliding with the monopole should satisfy the charge non conserved boundary condition. 
It apparently seems that the charge is not conserved in the fermion scattering with the GUT monopole.
But,
it was shown\cite{rubakov,callan,ezawa} 
that even if we take the charge non conserved boundary condition, the charge is conserved
when we take account of the quantum effects of the U(1) gauge fields. This is because
Coulomb repulsion around $r=0$ expels incoming fermions from the monopole.
Physically, the dyon ( monopole carrying U(1) charge ) has much large charging energy so that
the low energy fermion scattering with the monopole can not deposit its charge on the monopole.   
Consequently, the local U(1) gauge symmetry is not broken. Indeed,
the fermion-monopole system was solved exactly in a simplified model where only S wave components of
both fermions and the gauge fields are taken into account. The chiral anomaly was properly taken into account. 
Because the charge is conserved, the chirality must change when the fermions collide with the monopole.
Detail analysis has shown that the chirality non conservation can arise owing to the presence of the chiral condensate around the monopole.
That is, $\langle\bar{\psi}\psi\rangle\propto 1/r^3$.
The condensate results from the chiral anomaly, which is effective around the monopole
when charged fermions approach the monopole; $\partial_{\mu} J^{\mu}_5\propto \vec{E}\cdot\vec{B} \sim 1/ r^4$
with $\vec{E}\sim e\vec{r}/r^3$ and $\vec{B}\sim \vec{r}/(e r^3)$.
This fact is a byproduct of the analysis of the Rubakov effect.
( We can show  
that the condensates with baryon number but with no U(1) charge are present around a GUT monopole.
Although the boundary condition breaks both the charge and baryon number conservation,
only the non conservation of the baryon number remains to be effective.
This is the origin of the Rubakov effect. )

\vspace{0.1cm}
Now we explain the chiral condensate around the QCD monopole.
According to the assumption of the Abelian dominance, QCD monopoles and the Maximal Abelian gauge fields
are relevant to explain low energy phenomena such as the quark confinement.
There is a similarity in the fermion-monopole dynamics between the GUT monopole and the QCD monopole.
The QCD monopoles carry color magnetic charges coupled with the Maximal Abelian gauge fields.
Off diagonal gluons are massive and irrelevant to the low energy QCD physics. The similarity to the case of the Rubakov effect is obvious. 
In QCD, we have three types of monopoles, which are characterized by root vectors of SU(3),
 $\vec{\epsilon}_1=(1,0), \vec{\epsilon}_2=(-1/2,-\sqrt{3}/2)$ and $\vec{\epsilon}_3=(-1/2,\sqrt{3}/2)$.
They describe the couplings with the maximal Abelian gauge fields, $A_{\mu}^{3,8}$ such as $\epsilon_i^a A_{\mu}^a$.
For example a monopole with $\vec{\epsilon}_1$ couples only with $A_{\mu}^3=\epsilon_1^a A_{\mu}^a$. Thus, the quarks coupled with the monopole
are a doublet $q=(q^+,q^-,0)$ of the color triplet. Similarly the other monopoles couple with the quark doublets, $q=(q^+,0,q^-)$ and $q=(0,q^+,q^-)$. 
Thus, the analysis of the Rubakov effect is applicable to QCD under the assumption of the Abelian dominance.
Only a difference is that we have no appropriate way of choosing the charge non conserved boundary condition. 
But, even if we choose any boundary conditions with charge non conservation or chirality non conservation,
we find the presence of the chiral condensates around the monopoles.
As we explained above, the low energy quark scattering with the monopole can not change its color by emitting 
off diagonal gluons because they are massive. 
Thus, the charge conserved but chirality non conserved boundary condition must be the plausible boundary condition.


Indeed,
it is obvious that if the charge conserved but chirality non conserved boundary condition, 
$\psi_{+}(r=0)=\bar{\gamma}_0\psi_{+}(r=0)$ is taken,
the boundary condition itself causes the chiral condensate; 
there are no mechanism restoring the conservation of the chirality. 
The results are identical to the ones obtained by the use of the baryon number non conserved boundary condition, which gives rise to
the baryon number condensate around the GUT monopole. This is the essence of the Rubakov effect in the GUT monopoles.
Therefore, any choice of the charge non conserved or chirality non conserved boundary conditions 
gives rise to the formation of the chiral condensate around the QCD monopole.
As a consequence, the monopole condensation leads to the chiral condensation in vacuum.

Our result is based on the assumption of the Abelian dominance, which have been
shown to be valid in lattice gauge theories by using Maximal Abelian gauge.
Although the intimate relation between the monopole and the chiral symmetry breaking
depends on the specific gauge, 
recent works\cite{hasegawa} indicate the validity of the relation without the use of the Maximal Abelian gauge. 

We would like to comment that the assumption of the Abelian dominance is not applicable to the fermions in the adjoint representation of SU(3).
Actually, the naive application\cite{adj} of the Abelian dominance leads to non confinement of the fermions with no color charges of the gauge fields $A_{\mu}^{3,8}$.
For example, when the gauge group is SU(2), there are neutral components of the triplet fermions associated with the gauge fields $A_{\mu}^3$.
They are not confined if we naively apply the Abelian dominance.  
Hence, the analogy of the Rubakov effect is not applicable to the fermions in the adjoint representation.
The monopole do not necessarily carry the chiral condensate of the adjoint fermions.

\vspace{0.3cm}
We proceed to present a phenomenological model which realizes our result that the monopole condensation causes the chiral symmetry breaking.
For the purpose, we take a simple linear sigma model with $\rm SU(2)_L\times SU(2)_R$ symmetry.
The linear sigma model describes the spontaneous breakdown of the chiral symmetry.
We also take a simple model of SU(3) dual superconductor which describes the confinement by the monopole condensation.
The monopole fields are minimally coupled with the dual gauge fields whose fluxes are squeezed by the monopole condensation.
Then, the strings of color flux are formed to confine the quarks.
( We have three types of the monopoles $\Phi_i$ ( $i=1\sim 3$ ) and two dual gauge potentials $B_{\mu}^a$ ( $a=3,8$ )
in the SU(3) dual superconducting model. )
We introduce the monopole sigma meson ( $\sigma$ ) couplings with $\rm SU(2)_L\times SU(2)_R$ symmetry so that
a chiral condensate $\langle \sigma \rangle \propto 1/r$ is formed around a monopole.
We implicitly assume in the model that the sigma meson describes $\bar{q}q$.
With this chiral $\rm SU(2)_L\times SU(2)_R$ invariant coupling, both the monopoles and dual gauge fields couple with observed hadrons in the sigma model.
It apparently seems that both of the monopoles and dual gauge fields are observable.
But, as we show below, a Weyl symmetry in the dual superconducting model dictates that 
only a monopole excitation in the confining vacuum 
is observable.

\vspace{0.1cm}
Here we should make a comment that the physical observables in the SU(3) dual superconductor must be color SU(3) singlet.
A discrete symmetry of color SU(3) 
still remains in the dual superconducting model. It is a Weyl symmetry.
The physical observable should be invariant under the discrete symmetry.
The symmetry requires the system invariant under the three types of 
transformation of the dual gauge fields $B^3_{\mu}$ and $B^8_{\mu}$,

\begin{eqnarray}
\label{6}
(&B^3_{\mu}& \to -B^3_{\mu}, \,\,\,B^8_{\mu}\to B^8_{\mu} \quad),\quad 
(\quad B^3_{\mu} \to \frac{1}{2}B^3_{\mu}-\frac{\sqrt{3}}{2}B^8_{\mu}, \,\,\,B^8_{\mu}\to
-\frac{\sqrt{3}}{2}B^3_{\mu}-\frac{1}{2}B^8_{\mu}\quad) \nonumber \\
(&B^3_{\mu}& \to \frac{1}{2}B^3_{\mu}+\frac{\sqrt{3}}{2}B^8_{\mu},\,\,\,
B^8_{\mu}\to \frac{\sqrt{3}}{2}B^3_{\mu}-\frac{1}{2}B^8_{\mu}\quad),
\end{eqnarray}
which can be obtained by noting that the three types of the colors in quarks $q_1=(1,0,0)$, $q_2=(0,1,0)$ and $q_3=(0,0,1)$ are
transformed into each other under the elements $U_a$ of SU(3), $U_{(1,2)}q_1=q_2$, $U_{(1,3)}q_1=q_3$, and $U_{(2,3)}q_2=q_3$.
Using the elements, $B^3_{\mu}\lambda_3+B^8_{\mu}\lambda_8$ is transformed into 
$U_a^{\dagger}(B^3_{\mu}\lambda_3+B^8_{\mu}\lambda_8)U_a=B'^3_{\mu}\lambda_3+B'^8_{\mu}\lambda_8$;
$\lambda^3$ and $\lambda^8$ denote diagonal Gell-Man matrices. 
That is, we obtain the transformation, $B^3_{\mu}\to B'^3_{\mu}$ and $B^8_{\mu}\to B'^8_{\mu}$ which
are shown explicitly in eq(\ref{6}). The dual gauge fields $B^a_{\mu}$ minimally couple with the monopoles $\Phi_i$ 
such as $D^i_{\mu}\Phi_i=(\partial_{\mu}-ig_m\vec{\epsilon}_i\cdot\vec{B}_{\mu})\Phi_i $.
Hence,
the Weyl invariance requires that the monopole fields are transformed under the transformation in eq(\ref{6})
in the following,

\begin{equation}
\label{7}
(\,\, \Phi_1\to \Phi^{\dagger}_1,\,\,\,\Phi_2\to \Phi^{\dagger}_3,\,\,\, \Phi_3\to \Phi^{\dagger}_2\,\, ),\,\,
(\,\, \Phi_1\to \Phi^{\dagger}_3,\,\,\,\Phi_2\to \Phi^{\dagger}_2,\,\,\, \Phi_3\to \Phi^{\dagger}_1\,\, ),\,\,
(\,\, \Phi_1\to \Phi^{\dagger}_2,\,\,\,\Phi_2\to \Phi^{\dagger}_1,\,\,\, \Phi_3\to \Phi^{\dagger}_3\,\, )
\end{equation}
respectively. We find that the magnetic charge $\rho_m=\sum_{j=1,2,3}\Phi^{\dagger}_j(i\partial_0-g_m\vec{\epsilon}_j\cdot \vec{B}_0)\Phi_j+c.c.$
changes its sign under the transformation in eq(\ref{6}) and eq(\ref{7}); $\rho_m\to -\rho_m$.
Thus, the magnetic charge is not observable.

The Weyl invariance of the system shows that the color singlet physical states
are invariant under the transformation shown in eq(\ref{6}) and eq(\ref{7}). 
Therefore, the observable dual gauge fields are composites such as $(B^3_{\mu} )^2+(B^8_{\mu})^2$.
There are no single dual gauge fields $\sum_{a=3,8} C_a B^a_{\mu}$ with numerical coefficients $C_a$.
Similarly, the observable monopoles are composites with no magnetic charges such as $\sum_{i=1,2,3} |\Phi_i|^2$
in the phase with no monopole condensation $\langle \Phi_i \rangle=0$.
On the other hand, there is an observable state $\sum_{i=1,2,3}\delta\Phi_i$
in the phase with monopole condensation $\langle\Phi_i\rangle=v\neq 0$
because we can take the fields $\delta\Phi_i$ real with $\Phi_i=v+\delta\Phi_i$.  
Therefore, we expect that the monopole excitation can be identified with
an observed hadron.

\vspace{0.1cm}
In the present paper,
we tentatively take
a simple phenomenological model realizing the intimate relation mentioned above between the confinement and the chiral symmetry breaking. 
As a chiral model, 
we take a SU(2) linear sigma model
of the field $\Sigma\equiv \sigma +i\vec{\pi}\cdot\vec{\tau}$.
The chiral condensate is described by the non zero expectation value of $\sigma$; $\langle\sigma\rangle\neq 0$.
In order for the magnetic charges to be sources of the sigma field, we introduce the interaction 
$h\sigma \sum_i|\Phi_i|^2$
between the monopoles and sigma
field.
The term gives rise to the sigma field $\sigma \sim 1/r$ for a point-like monopole $\sum_i|\Phi_i|^2\sim \delta^3(x)$. 
Thus, we assume the interaction between the monopole fields $\Phi_i$ and chiral field $\Sigma$

\begin{equation}
L_{\rm{int}}=h\sqrt{\frac{\rm{Tr}(\Sigma^{\dagger}\Sigma)}{2}}\sum_i|\Phi_i|^2=h\sqrt{\sigma^2+\vec{\pi}^2}\sum_i|\Phi_i|^2
\end{equation}
with $h>0$,
where $\rm{Tr}(\Sigma^{\dagger}\Sigma)=2(\sigma^2+\vec{\pi}^2)$.
The interaction is invariant under the $\rm SU(2)_L\times SU(2)_R$ transformation $\Sigma\to U_{L}^{\dagger}\Sigma U_{R}$
with $U_{L,R}$ elements of SU(2). It is also invariant under the transformation of the Weyl symmetry.
We should stress that the interaction term between the monopole and the chiral field is determined without
any ambiguity. The term is taken such that it gives rise to the chiral condensate locally around a monopole.

A model of a dual superconductor is taken as,

\begin{equation}
L_{\Phi}=\sum_{i=1\sim 3}(\frac{1}{2} |D_{\mu}^i\Phi_i|^2+\mu^2|\Phi_i|^2-\lambda' |\Phi_i|^4)-\frac{1}{4}((B^3_{\mu\nu})^2+(B^8_{\mu\nu})^2)
\end{equation}
with $B^a_{\mu\nu}=\partial_{\mu}B^a_{\nu}-\partial_{\nu}B^a_{\mu}$ and $\lambda'>0$,
where we have denoted
three types of the monopoles $\Phi_i$ and two types of dual gauge fields $B_{\mu}^a$ ( $a=3,8$ );
$D_{\mu}^i\equiv \partial_{\mu}+ig_m\epsilon_i^a B_{\mu}^a$ 
with root vectors of SU(3), $\vec{\epsilon}_1=(1,0), \vec{\epsilon}_2=(-1/2,-\sqrt{3}/2)$ and $\vec{\epsilon}_3=(-1/2,\sqrt{3}/2)$.
We denote the magnetic charge as $g_m=4\pi/g$ ( $g$ is the gauge coupling constant of QCD ).      

On the other hand,
a sigma model is taken such that,

\begin{equation}
L_{\Sigma}=\frac{1}{4}\rm{Tr}(\partial_{\mu}\Sigma^{\dagger}\partial^{\mu}\Sigma)-\frac{m^2}{4}\rm{Tr}(\Sigma^{\dagger}\Sigma)
-\frac{\lambda}{4} (\rm{Tr}(\Sigma^{\dagger}\Sigma))^2
=\frac{1}{2}(\partial\sigma)^2+\frac{1}{2}(\partial\vec{\pi})^2-\frac{m^2}{2}(\sigma^2+\vec{\pi}^2)-\lambda(\sigma^2+\vec{\pi}^2)^2,
\end{equation}
with $\lambda>0$,
where we have a real mass $\rm{m}$ contrary to the ordinary sigma model in which 
the mass is taken to be imaginary for the realization of the spontaneous chiral symmetry breaking.
But, the chiral symmetry breaking is caused by the condensation of the monopoles in our model.
The total Lagrangian is given by $L_{\Phi}+L_{\Sigma}+L_{\rm int}$.

\vspace{0.1cm}
In order to find the vacuum configurations of $\sigma$ and $\Phi_i$
 with $\vec{\pi}=0$ and $B_{\mu}^a=0$, we derive the equations of motion of the sigma $\sigma $ and the monopoles $\Phi_i$ homogeneous in space-time,

\begin{eqnarray} 
\label{9}
m^2\sigma+4\lambda \sigma^3-h\sum_{i=1,2,3}|\Phi_i|^2=0 \\
(-\mu^2+2\lambda'|\Phi_i|^2-h\sigma )\Phi_i=0 \quad \mbox{for} \quad i=1,2,3
\label{10}
\end{eqnarray}

The non trivial solutions are given by

\begin{equation}
\label{11}
|\Phi_0|^2=\frac{\mu^2+h\sigma_0}{2\lambda'} \quad \mbox{and} \quad 4\lambda\sigma_0^3+m^2\sigma_0=\frac{3h(\mu^2+h\sigma_0)}{2\lambda'}
\end{equation}
where we put $\Phi_i\equiv \Phi_0$ for all $i$. The potential of the solutions is
given by

\begin{equation}
V(\sigma_0, \Phi_0)=-3\mu^2|\Phi_0|^2+3\lambda'|\Phi_0|^4+\frac{1}{2}m^2\sigma_0^2+\lambda\sigma_0^4-3h\sigma_0|\Phi_0|^2
=-\frac{3\mu^2}{2}|\Phi_0|^2-\lambda\sigma_0^4
\end{equation}
where we used the equations (\ref{11}) to rewrite the potential $V$.

There is a trivial solution $\Phi_0=0$ and $\sigma_0=0$ in eq(\ref{9}) and eq(\ref{10}).
Obviously, the solutions representing the condensates $\sigma_0\neq 0 \neq\Phi_0$ have lower energies than  the energy of the trivial solution.
Therefore, we find that there are condensates $\sigma_0\neq 0$ and $\Phi\neq 0$ in the ground state.
We note that when $\Phi_0\neq 0$, we obtain $\sigma_0\neq 0$.
Namely, the monopole condensation causes the chiral symmetry breaking.
We also note that when the monopole condensate disappears $\Phi_0\to 0$, the chiral condensate also disappears.
It implies the fact that the deconfinement and the chiral symmetry restoration occur simultaneously,
although we need to examine thermodynamical potentials to confirm it.  

We make a comment that even if the monopole condensates take place, 
the chiral condensate does not arise when we switch off the coupling, i.e. $h=0$ between the monopoles and the chiral field. 
Thus, the coupling is essential for the chiral condensate.
Therefore, the model describes the chiral symmetry breaking induced by the monopole condensation.

\vspace{0.1cm}
We calculate the masses of the particles in the model. For the purpose, we put $\sigma=\sigma_0+\delta\sigma$ and $\Phi_i=\Phi_0+\delta\Phi_i$
in the Lagrangian  $L_{\Phi}+L_{\Sigma}+L_{\rm int}$ and extract the quadratic terms in $B_{\mu}^a$, $\delta\sigma$ and $\delta\Phi_i$,

\begin{eqnarray}
&&\sum_{i=1\sim 3}\frac{1}{2}|\partial_{\mu}\delta\Phi_i|^2+\frac{1}{2}(\partial_{\mu}\delta\sigma)^2 -\frac{1}{4}((B_{\mu\nu}^3)^2+(B_{\mu\nu}^8)^2)  \nonumber \\
&+&\sum_{i=1\sim 3}(g_m\vec{\epsilon}_i\cdot\vec{B}_{\mu}\Phi_0)^2-\Big(\sum_{i=1\sim 3}(2\mu^2+2h\sigma_0)(\delta\Phi_i)^2+6\lambda\sigma_0^2(\delta\sigma)^2
-2\sum_{i=1\sim 3} h\Phi_0\delta\sigma\delta\Phi_i+\frac{m^2}{2}(\delta\sigma)^2\Big) \nonumber \\
&=& \sum_{i=1\sim 3}\frac{1}{2}(\partial_{\mu}\delta\Phi'_i)^2+\frac{1}{2}(\partial_{\mu}\delta\sigma')^2-\frac{1}{4}((B_{\mu\nu}^3)^2+(B_{\mu\nu}^8)^2) \nonumber \\
&+&g_m^2\frac{3\Phi_0^2}{4} \Big( (B_{\mu}^3)^2+(B_{\mu}^8)^2\Big)-\frac{M^2}{2}\Big((\delta\Phi_1')^2+(\delta\Phi_2')^2\Big)-\frac{M'^2}{2}(\delta\Phi_3')^2-\frac{M_{\sigma}^2}{2}
(\delta\sigma')^2
\end{eqnarray}

with

\begin{eqnarray}
M^2&\equiv&4(\mu^2+h\sigma_0), \quad M'^2\equiv2\Big(\mu^2+h\sigma_0+3\lambda\sigma_0^2+\frac{m^2}{4}+\sqrt{(\mu^2+h\sigma_0-3\lambda\sigma_0^2-\frac{m^2}{4})^2
+3h^2\Phi_0^2} \,\Big) \\
M_{\sigma}^2&\equiv&2\Big(\mu^2+h\sigma_0+3\lambda\sigma_0^2+\frac{m^2}{4}-\sqrt{(\mu^2+h\sigma_0-3\lambda\sigma_0^2-\frac{m^2}{4})^2
+3h^2\Phi_0^2} \,\Big)
\end{eqnarray}
where the fields $\delta\Phi_i'$ and $\delta\sigma'$ are defined in the following,

\begin{eqnarray}
\delta\Phi_1'&=&\frac{1}{\sqrt{2}}(\delta\Phi_1-\delta\Phi_2), \quad  \delta\Phi_2'=\frac{1}{\sqrt{6}}(\delta\Phi_1+\delta\Phi_2-2\delta\Phi_3) \\
\delta\Phi_3'&=&\frac{1}{\sqrt{x_+}}(\delta\Phi_1+\delta\Phi_2+\delta\Phi_3+\frac{0.5(M^2-M'^2)}{h\Phi_0}\delta\sigma) \\
\delta\sigma'&=&\frac{1}{\sqrt{x_-}}(\delta\Phi_1+\delta\Phi_2+\delta\Phi_3+\frac{0.5(M^2-M_{\sigma}^2)}{h\Phi_0}\delta\sigma)
\end{eqnarray}

with 

\begin{equation}
x_+\equiv\frac{3(M'^2-M_{\sigma}^2)}{M^2-M_{\sigma}^2}, \quad  x_-\equiv\frac{3(M'^2-M_{\sigma}^2)}{M'^2-M^2}.
\end{equation}
We can show that the parameters $M_{\sigma}^2$, $x_+$ and $x_-$ are non negative.


\vspace{0.1cm}
We find that the mass of the monopoles $\delta\Phi_1'$ and $\delta\Phi_2'$ is given by $M$, 
while the mass of the monopole $\delta\Phi_3'$ is $M'$, which is heavier than the mass $M_{\sigma}$ of
the sigma meson $\delta\sigma'$; $M'>M_{\sigma}$.
The dual gauge fields $B_{\mu}^3$ and $B_{\mu}^8$ are degenerate; their mass $M_B$ is given such that $M_B^2=3g_m^2\Phi_0^2/2$.
The masses of pion $\vec{\pi}$ vanish because they are Nambu-Goldstone bosons associated with the chiral symmetry.
Furthermore, in the Bogomol'ny limit, i.e. $\lambda'=3g_m^2/16$, the mass $M_{B}$ is equal to $M$.
( It has been discussed that the relation  $\lambda'\simeq 3g_m^2/16$ may hold in real QCD ).
In general there is the mass hierarchy such that $M'>M>M_{\sigma}$.

As we mentioned, the physical particles in the dual color superconducting model must be symmetric under the Weyl symmetry.
Therefore,
we find that the physical particles are the monopole $\delta \Phi'_3$ with mass $M'$, sigma meson $\delta \sigma$ with mass $M_{\sigma}$,
and massless pion $\vec{\pi}$; $M'>M_{\sigma}$. The dual gauge fields are not observable, although their composites such as 
$(B^2_{\mu})^2+(B^8_{\mu})^2$ are observable. In simple dual superconducting models not involving hadrons previously discussed, 
the monopoles play only a role of the confinement. But 
the monopole $\delta \Phi'_3$ is observable in our model.


\vspace{0.1cm}

It is easy to see how 
the monopole couples with ordinary hadrons, e.g. pion.
Their coupling is described by $\sum_i h\sqrt{(\sigma_0+\delta\sigma)^2+\vec{\pi}^2}\,(\Phi_0+\delta\Phi_i)^2$.
In particular, three point interactions between pions and the monopole 
are given by

\begin{eqnarray}
\label{three}
&&\quad \quad \quad \frac{h\Phi_0\vec{\pi}^2}{\sigma_0}\sum_{i=1\sim 3}\delta\Phi_i-\frac{3h\sigma_0\Phi_0^2}{2}\,\frac{\delta\sigma}{\sigma_0}\frac{\vec{\pi}^2}{\sigma_0^2} \nonumber \\
&=&
\frac{\sqrt{3}h\Phi_0\vec{\pi}^2}{\sigma_0}\Big(\sqrt{\frac{M_{\sigma}^2-M^2}{M_{\sigma}^2-M'^2}}\delta\Phi_3'+\sqrt{\frac{M^2-M'^2}{M_{\sigma}^2-M'^2}}\delta\sigma'\Big) \nonumber \\
&&-\frac{3h\Phi_0^2\vec{\pi}^2}{2\sigma_0^2}\Big(-\sqrt{\frac{M'^2-M^2}{M'^2-M_{\sigma}^2}}\delta\Phi_3'+
\sqrt{\frac{M^2-M_{\sigma}^2}{M'^2-M_{\sigma}^2}}\delta\sigma' \Big).
\end{eqnarray}
The interaction represents the decay of the physical monopole $\delta\Phi_3'$ and sigma meson $\delta\sigma'$ into pions.
On the other hand,  
there are no three points interactions between pions and the unphysical monopoles $\delta\Phi_{1,2}'$.

Because the monopoles $\delta\Phi_i'$ are scalar and isoscalar, their quantum number is given by $J^{PC}=0^{++}$.  
Examining the three point interactions in eq(\ref{three}), the observable
monopole $\delta \Phi'_3$ could be identified
as hadrons $f_0$. 
The physical parameters in our model may be determined using the identification of the monopole as an observed hadron
and comparing string tension of flux tubes in our model with the observed string tension, i.e. Regge slope.
We can also use the strength of the decay modes of the monopole and sigma meson into pions for the
determination.

%

\vspace{0.2cm}

We have discussed the presence of the chiral condensate around a QCD monopole by using the analogy between GUT monopole in
the Rubakov effect and QCD monopole. The analogy is obvious when we assume
the Abelian dominance, in which the relevant components to the study of
the chiral condensate around a QCD monopole are the maximal Abelian gauge fields and massless quarks as well as the QCD monopoles;
off diagonal components of the SU(3) gauge fields are massive and irrelevant. 
It follows from the chiral condensate around a QCD monopole that 
the monopole condensation causes the chiral symmetry breaking as well as the confinement.
To make a phenomenological model in which the confinement causes the chiral symmetry breaking,
we have taken a linear sigma model coupled with the QCD monopoles $\Phi$.
It is interesting that the monopole $\delta\Phi'_3$ is observable 
through their coupling with ordinary hadrons. On this point, we point out\cite{heavy} that QCD monopoles play a role in the production
of quark gluon plasma
in high energy heavy ion collisions.
There are several chiral models in which the chiral symmetry is spontaneously broken.
In order to find a realistic chiral model, we need to examine
the models improved such that the chiral symmetry breaking is induced by the monopole condensation.


 \vspace{0.2cm}
The author
expresses thanks to members of KEK for useful comments
and discussions.



\end{document}